\documentclass[10pt,twocolumn,letterpaper]{article}

\usepackage{cvpr}              

\usepackage{graphicx}
\usepackage{amsmath}
\usepackage{amssymb}
\usepackage{float}
\usepackage{amsmath}
\usepackage{algpseudocode}
\usepackage{algorithm}
\usepackage{multirow}

\usepackage{booktabs}
\usepackage{enumitem} 
\usepackage{comment}

\usepackage[pagebackref,breaklinks,colorlinks]{hyperref}

\usepackage[capitalize]{cleveref}
\crefname{section}{Sec.}{Secs.}
\Crefname{section}{Section}{Sections}
\Crefname{table}{Table}{Tables}
\crefname{table}{Tab.}{Tabs.}

\def\cvprPaperID{11678} 
\def\confName{CVPR}
\def\confYear{2022}

\begin{document}

\title{Differentiable Microscopy for Content and Task
Aware Compressive Fluorescence Imaging}

\author{Udith Haputhanthri $^1$, Andrew Seeber $^1$, Dushan Wadduwage$^{1,*}$\\ \\
$^{1}$Center for Advanced Imaging, Faculty of Arts and Sciences, Harvard University, Cambridge, USA
\\
{\tt\small $^{*}$wadduwage@fas.harvard.edu}}

\maketitle

\begin{abstract}
The trade-off between throughput and image quality is an inherent challenge in microscopy. To improve throughput, compressive imaging under-samples image signals; the images are then computationally reconstructed by solving a regularized inverse problem. Compared to traditional regularizers, Deep Learning based methods have achieved greater success in compression and image quality. However, the information loss in the acquisition process sets the compression bounds. Further improvement in compression, without compromising the reconstruction quality is thus a challenge. In this work, we propose differentiable compressive fluorescence microscopy ($\partial \mu$) which includes a realistic generalizable forward model with learnable-physical parameters (e.g. illumination patterns), and a novel physics-inspired inverse model. The cascaded model is end-to-end differentiable and can learn optimal compressive sampling schemes through training data. With our model, we performed thousands of numerical experiments on various compressive microscope configurations. We show that learned sampling encodes important information about the specimens in the illumination field of the microscope allowing higher compression up to $\times 1024$. We further utilize our framework for Task Aware Compression. The experimental results show superior performance on the cell segmentation task.

\end{abstract}

\section{Introduction}


High-throughput high-content microscopy is an essential tool in today's biology and medicine. With applications ranging from diagnostics and drug discovery to neuroscience, gigabyte-scale microscopy image datasets are routinely collected, stored, and analyzed. To collect these large datasets, much progress has been made in fast imaging systems with newer optical designs, better mechanical movements, and faster and more multiplexed imaging sensors. Despite these significant advances, collecting terabyte-scale data is challenging. For instance, tissue-clearing techniques can render the whole mouse central nervous system transparent; yet imaging the whole mouse brain at diffraction-limited resolution is not possible even within weeks. Therefore, the resolution is often sacrificed in favor of scale and time or vice versa. However, the recent deep learning revolution may present a new opportunity to overcome this fundamental challenge through compressive imaging.


Compressive imaging frameworks can generally be modeled with a forward model and an inverse model \cite{2016nips_chakrabarti,2017dirtypixels_diamond,2019opticsexpress_hershko,2019iccp_kellman,2019optica_cheng,2021iccv_cooke}. The forward model represents an optical system that forms images starting from the illumination and ending at the photodetection. The inverse model is a restoration framework where the original image of the biological specimen is restored from the detected measurements. To compressively image, the forward model maps the original image signal to a lower-dimensional measurement space. The dimensionality reduction, or the compression, is proportional to the improvement in imaging speed. However, compression makes the inverse problem ill-posed, i.e. same measurements can map to multiple image reconstructions \cite{Pronina2020MicroscopyIR}. To solve the ill-posed inverse problem, compressive sensing mechanisms with   sparsity-based regularizers  \cite{comp_sampling1,comp_sampling2,comp_sampling3,comp_sampling4} were traditionally used. Today, however, deep learning with reconstruction objectives, has replaced many traditional inverse solvers \cite{restoration1,restoration2,restoration3,restoration4}. One can either repurpose current image-to-image translation methods \cite{pix2pix,swinir} or develop new deep-learning pipelines on detected image measurements to restore the underline images of the biological specimens. Nevertheless, the restoration capabilities of these algorithms depend on the compressive sampling scheme, i.e. how the forward model reduces the measurement dimensionality. 

Traditionally, compressive imagers perform incoherent random sampling, or engineered sampling from orthogonal bases (such as the Hadamard basis). In this work, we hypothesize that better compression schemes can be learned through data and physics-based models. To this end, we propose an end-to-end differentiable compressive microscope with a generalizable physics-constrained forward model and a physics-inspired inverse model. The proposed framework allows tuning physical parameters (e.g. illumination patterns) of the optical forward model constrained to physical limitations. The method essentially learns optimal sampling at a given compression level for training data. To overcome the discrepancy between the framework and the optical setup, we propose a novel differentiable noise model which mimics the major sources of noise: content-dependent Poisson noise and read noise. This allows the forward and inverse models to learn the physical noise statistics. We demonstrate the performance of our method by reconstructing images of biological specimens at high compression levels outperforming traditional sampling schemes. Our contributions are: 

\begin{itemize}
\item End-to-end differentiable fluorescence microscopy to learn free parameters of microscopes through classical backpropagation.

\item Demonstration of the proposed microscopy method on illumination pattern learning. 

\item A frequency-domain optimization technique  to learn illumination patterns with improved end-to-end joint optimization stability. 

\item A locality-aware physics-motivated upsampling block to effectively revert missing information: We discuss the superiority of the proposed upsampling block with respect to the classical transpose convolution. We also experimentally show that one can fuse the proposed upsampling block with state-of-the-art super-resolution pipelines to achieve high-resolution reconstruction quality.

\item A differentiable stochastic noise model that mimics content-dependent Poisson noise and detector-dependent read noise.

\item Demonstration of the superiority of the proposed microscopy framework for content-aware compressive sampling. We compare the proposed framework with traditional compressive sampling methods including pseudo-random, engineered (from a Hadamard basis), and uniform. 

\item Demonstration of the superiority of the proposed microscopy framework for task-aware sampling on \textit{segmentation}.



\end{itemize}

\section{Related Work}

Deep learning-powered computational imaging to improve imaging hardware has been explored by a few. Chakrabarti et al. \cite{2016nips_chakrabarti} optimized a camera sensor's color multiplexing pattern for grayscale sensor parallelly to an inverse reconstruction model. Diamond et al. \cite{2017dirtypixels_diamond} proposed an algorithm for context-aware denoising and deblurring that can be jointly optimized with the inverse model for cameras. Diffractive Optical Elements (DOEs) in the camera forward model are also widely explored in the recent literature. Work done by Sitzmann et al. \cite{sitz2018} explored a wave-based image formation model using diffraction as the forward model. Peng et al. \cite{peng2019} address high-resolution image acquisition using Single Photon Avalanch Photodiodes where the forward model is also based on diffractive optics-based learnable Point Spread Functions (PSF). Apart from those, several works targetted shift-invariant PSFs \cite{sun2020}, high-dynamic-range imaging \cite{sun2020cvpr} through diffractive optics as well. Hougne et al. \cite{2020as_hougne} proposed a reconfigurable metamaterial-based device that can emit microwave patterns capable of encoding task-specific information for classification.

Recently, learning-based computational microscopy approaches have also been proposed to optimize the forward optics model (i.e. the mathematical model of the microscope) jointly with the inverse model (i.e. the image reconstruction algorithm). These works are largely on optimization of LED illumination of microscopes, with the exception of finding emitter color when using a grayscale camera in fluorescence microscopy by Hershko et al. \cite{2019opticsexpress_hershko}.

Several groups optimized the LED patterns on various types of microscopes. Work done by \cite{2019iccp_kellman,2019optica_cheng} has increased the temporal resolution in Fourier Ptychography Imaging by incorporating illumination optimization. Recent work done by Cooke et al. \cite{2021iccv_cooke} illustrates the use of pattern optimization in virtual fluorescence microscopy where the inverse model converts the acquired images from unlabelled samples to labeled fluorescence images. They have shown the superiority of their method compared to virtual fluorescence microscopy with a fixed optical system. The previous work has utilized illumination pattern optimization for phase information extraction as well \cite{2018plos_diederich,2019tcm_kellman}. Apart from that, the works done by \cite{2017teachmic_horstmeyer,2019boe_muthumbi,2020opticsletters_kim,2021icassp_cooke} have shown the improved performance via optimization of LED illumination patterns for malaria-infected cell classification in cell imaging.

All illumination pattern optimization methods in the literature on microscopy focused on LED array illuminations. They considered individual LED sub-illuminations while assuming a linear forward model without noise or with only \textit{independent and identically distributed (iid)} noise that violates experimental situations. Limiting to a single microscopy structure (e.g. LED array illumination microscopy) is also limiting the generalizability of existing methods on other microscope modalities. The recent work proposed by Sun et al. \cite{2020blackhole_sun} illustrates an optimized sensor design for better reconstruction with a probabilistic sensor sampling strategy while demonstrating the importance of realistic noise modeling compared to the previous illumination optimization methods.

In contrast to the existing methods, we propose a unified generalized framework to jointly optimize the forward and inverse model of fluorescence microscopy. To this end, we demonstrate the achievable higher compression through the proposed framework by optimizing the excitation patterns of the microscopy. Even though we have not demonstrated in this paper, one can also incorporate learnable diffractive optical elements as well to the existing pipeline by simply making the PSFs learnable of the proposed framework. We further propose a differentiable noise model to mimic major noise sources while allowing end-to-end training. To the best of our knowledge, we are the first to focus on compressive sampling in microscopy through deep learning to improve throughput while keeping the reconstruction performance degradation minimal.

\begin{figure*}[hbt!]
\begin{center}
\includegraphics[width=1.9\columnwidth]{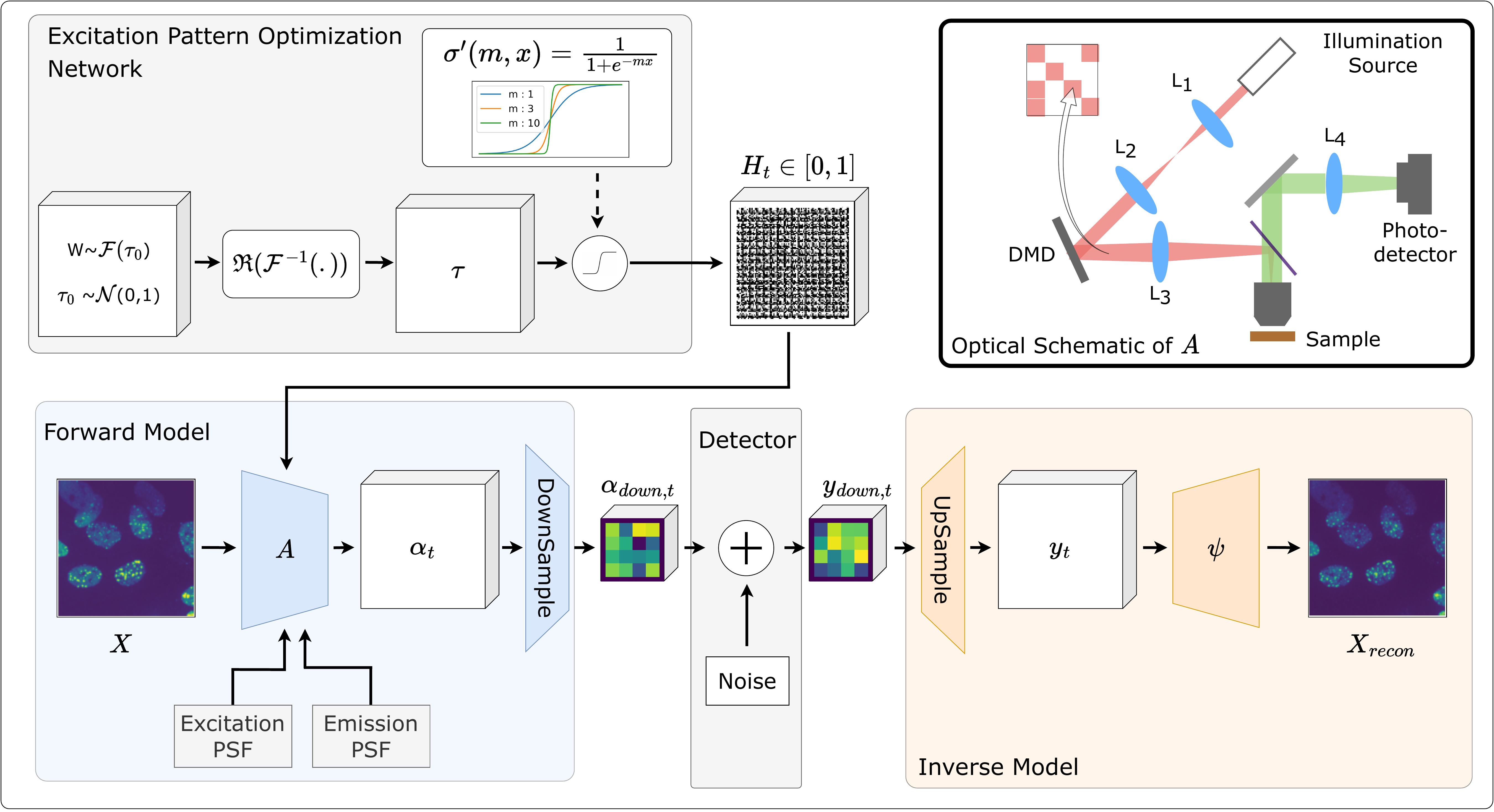}
\end{center}
\caption{End-to-end differentiable microscopy: The excitation pattern optimization network generates binary excitation patterns $H_t$. They are then utilized to image the biological samples where the optical setup is modeled through the forward model. The resultant light field is then demagnified optically and acquired through a photo-detector array. The inverse model super-resolves the acquired image and produces a high-resolution image of the sample. In the optical schematic of the Forward model (A), the illumination source illuminates the digital micromirror device (DMD) that introduces $H_t$. The patterns are projected onto the sample through the tube-lens L3 and microscope objective ($exPSF$). Then the emission light is collected through the same objective and relayed to the detector through the tube lens L4 ($emPSF$ and $demag(.)$).}
\label{fig:end2end}
\end{figure*}

\section{Preliminaries}
\subsection{Wide-field Fluorescence Microscopy}

The forward light propagation model of a wide-field fluorescence microscope in 2D can be described by the eq. \ref{eq:classical}. 

\begin{equation} 
\begin{split}
\alpha=\operatorname{emPSF} \ast X
\end{split}
\label{eq:classical}
\end{equation}

where, $\ast, \alpha, X, \operatorname{emPSF}$ are the convolution operation, acquired image, specimen (corresponding to the ground-truth image), emission point spread function (PSF) \cite{psfs}.

\subsection{De-scattering with Excitation Patterning (DEEP)}

De-scattering with Excitation Patterning (DEEP) is a structured illumination method recently proposed by Zheng et al. \cite{deeptfm} to image through scattering tissues. In their method, they proposed to use randomly initialized multiple excitation patterns to image deep through scattering tissues at high resolution. Usually, scattering acts as a low-pass filter blocking high spatial frequencies of the image. DEEP samples and retains high-frequency components (that are otherwise lost due to scattering) by encoding them onto the pass-band, using excitation patterns. We generalized their idea of the use of excitation patterns to encode high-frequency content to the passband of a low-pass filter. In our case, we demagnify the encoded image before detection, to reduce the dimensionality of measurements; this acts as our low-pass filter.


\section{Differentiable Compressive Microscopy}

In this work, we propose \textit{end-to-end differentiable microscopy} ($\partial \mu$) to optimize the physical parameters of a computational microscope for a specific goal. We demonstrate $\partial \mu$ on a general coded-aperture-based optical design similar to DEEP \cite{deeptfm}. See the optical schematic in Fig \ref{fig:end2end}. Briefly, a digital micro-mirror device (DMD) is placed on a conjugate image plane at the microscope's illumination beam path. Then a structured illumination, based on a predefined excitation pattern, is projected onto the specimen using the DMD. The  fluorescence signal is then imaged onto a multi-pixel photodetector. Based on the magnification, multiple DMD pixels can map onto each detector pixel for compressive sampling.
We note that most computational microscopes (such as the single-pixel compressive microscope and wide-field fluorescence microscope) are special cases of the proposed design.

Traditionally, a set of random or engineered excitation patterns on the DMD are used to encode signals to the detector pixels. In this work, we demonstrate that learning the excitation patterns using $\partial \mu$ is superior to that. To this end, we propose an end-to-end differentiable framework having a forward model and an inverse model. The differentiable forward model has three major components: 1) The excitation pattern optimization network, 2) The optical demagnification model, and 3) The differentiable photodetector model. The inverse model is a deep neural network that consists of an upsampling network and a reconstruction network. The overall framework is shown in Fig. \ref{fig:end2end}

In the following sections, we discuss the physics-based forward model (section \ref{sec:Physics-based_Forward_Model}), the differentiable implementation of the forward model (section \ref{sec:detector_model}), the inverse model (section \ref{sec:inverse_model}), and end-to-end training (section \ref{sec:end-to-end_training}).

\subsection{The Physics-based Forward Model}
\label{sec:Physics-based_Forward_Model}

Eq. \ref{eq:forward_our}, \ref{eq:forward_our2} show the mathematical model of the proposed microscope. 
 
\begin{equation} 
\label{eq:forward_our}
\begin{split}
\alpha_t(x, y)=\operatorname{emPSF}(x, y) \ast \Big\{(&\operatorname{exPSF}(x, y) \\ \ast H_{t}&(x, y)) \circ X(x, y)\Big\} 
\end{split}
\end{equation}

\begin{equation} 
\label{eq:forward_our2}
\alpha_{down, t}(x, y) = demag(\alpha_t(x, y))
\end{equation}

 Respectively, $H_{t}$, $X(x, y)$, $\alpha_t(x, y)$, $\alpha_{down, t}(x, y)$ represent the $t^{th}$  excitation pattern, the biological specimen, the pattern-encoded image before and after demagnification. $demag(.)$, and $\circ$ represent optical demagnification and pixel-wise multiplication. Unless otherwise specified the point spread functions, $\operatorname{emPSF}$ and $\operatorname{exPSF}$, are impulse kernels. $H_t$ is generated by the Digital Micromirror Device (DMD). The demagnified optical image, $\alpha_{down, t}(x, y)$, is acquired by the photo-detector array.

\subsection{The Differentiable Implementation of the Physics-based Forward Model}
\label{sec:detector_model}

The forward model should be differentiable to learn its physical parameters through backpropagation. Here we introduce a differentiable implementation of the forward model.
    \paragraph{Excitation pattern optimization network}
     Excitation patterns generated by the DMD are binary and hence the direct DMD model is not differentiable. We propose a new differentiable network model that approximates the real DMD. We also found that directly learning the patterns themselves is unstable. Thus we introduced a reparameterization network to represent parameters in the frequency domain. The proposed model (shown in Fig. \ref{fig:end2end}) consists of a custom sigmoid activation for binary pattern generation, frequency domain optimization, and Sigmoid-friendly initialization through Fast Fourier Transform.

    \begin{enumerate}
        \item We propose \textit{Custom sigmoid activation for binary pattern generation} to make the excitation patterns binary. 
        
        \begin{equation}
        \sigma^{\prime}(m, x)=\frac{1}{1+e^{-m x}}
        \label{eq:custom_sig}
        \end{equation}
        Proposed custom sigmoid activation is shown in equation \ref{eq:custom_sig}. The difference between conventional sigmoid activation \cite{nwankpa2018activation} and the custom sigmoid is the hyperparameter $m$. Throughout the learning of the end-to-end model, we increase $m$ gradually through a specific schedule. With this, we keep the values on the excitation patterns more towards {0, 1} without significant degradation in the performance.
        
        We find that using the schedule presented in the algorithm \ref{eq:m-schedule} is beneficial for the stability of the learning.  Refer to supplementary materials for further details regarding the schedule selection for hyper-parameter $m$.

        \begin{algorithm}
        \begin{algorithmic}
        \If{$epoch > epoch_{baseline} $}
            \State Optimize $H_t$ end-to-end with the inverse model
            \If{$epoch > epoch_{cutoff}$ and $epoch\% epoch step==0$} 
                \State $m=m+1$
            \Else
                \State $m=1$
            \EndIf 
        \Else   
            \State Optimize only the inverse model
        \EndIf 
        \end{algorithmic}
        \caption{Schedule for updating hyper-parameter $m$ in custom sigmoid (For experiments with U2OS Cell dataset in section. \ref{sec:datasets}, $epoch_{baseline}= 12150, epoch_{cutoff}= 18630, epoch step= 810$)}
        \label{eq:m-schedule}
        \end{algorithm}
        
        Equation \ref{eq:sig_out} shows the resultant excitation patterns. $\tau$ is the input which is discussed in the next section.
        
        \begin{equation}
        \label{eq:sig_out}
        H_t = \sigma' (m, \tau)
        \end{equation}
        
        \item \textit{Frequency domain optimization}: Even though the excitation patterns are in the spatial domain, through frequency domain optimization, we can learn weights not only considering the spatial features but also the frequency features. We found that incorporating frequency domain optimization results in more robust translation-invariant excitation patterns which can encode highly important features of the sample. Please refer to the section \ref{ablation_study} for further discussion along with the demonstration of the effectiveness of the frequency domain optimization.
        
        \begin{equation}
        \tau = \Re(\mathcal{F}^{-1}(W))
        \label{eq:ifft_out}
        \end{equation}
        
        Equation \ref{eq:ifft_out} shows the frequency domain weight optimization where $\mathcal{F}^{-1}(.)$, $\Re(.)$, and $W$ are respectively the Inverse Fourier Transform operation, the real part of a complex number, and the weights that are learned during the training.
        
        \item \textit{Sigmoid-friendly initialization}: The main challenge of incorporating frequency domain optimization is, it results in complex values that have vastly different ranges compared to their input range. Therefore feeding the output of that into the custom sigmoid leads to vanishing gradients \cite{Squartini2003_solvevanish}. To overcome this, we propose a weight initialization method through Fourier Transform on standard normal distributed noise (eq. \ref{eq:Ht_weight_init}). 
        
        \begin{equation}
        \begin{aligned}
        &\mathrm{W} \sim \mathrm{\mathcal{F}}\left(\tau_{0}\right) \\
        &\tau_{0} \sim \mathcal{N}(\mu= 0, \sigma = 1)
        \end{aligned}
        \label{eq:Ht_weight_init}
        \end{equation}

        Here $\mathcal{N}(\mu= 0, \sigma = 1)$ represents the standard normal distribution.

    \end{enumerate}
    
    \paragraph{Optical demagnification model}
    To mimic optical demagnification, we derive the \textit{sum pooling operation} through average pooling as shown in eq. \ref{eq:sumpool}
    
    \begin{equation}
    sum pool_{n\times n}(X) = Average Pool_{n\times n}(X) \times n^2 
    \label{eq:sumpool}
    \end{equation}
    
    where $n$ is the kernel size and $X$ is the input image.

        
    \paragraph{Differentiable Photo-detector model}
    The photon budget of the chosen fluorophore in fluorescence microscopy sets the upper bound to lighting conditions. To avoid photobleaching, many samples, such as voltage probes in the brain, must be imaged in low light. Therefore increasing the illumination intensity until getting a desirable SNR is not always an option. Thus a robust differentiable noise model is proposed to mimic real-world noise sources in the training process which ultimately makes the imaging procedure robust to noise. We consider the major noise components injected at the photodetector which are Poisson-distributed signal noise and read noise.
    
    The considered content-dependent Poisson noise and read noise addition are shown in eq. \ref{eg:pos_noise}.
    
    \begin{equation}
    y_{down, t} =  y^{poiss}_{down, t} + y^{normal}_{down, t} 
    \label{eg:pos_noise}
    \end{equation}
    where,
    
    \begin{equation}
    \begin{aligned}        
        y^{poiss}_{down, t} & \sim \operatorname{Poiss}(\alpha_{down, t})\\
        y^{normal}_{down, t} & \sim \mathcal{N}(\mu= 0, \sigma= \sigma _{read})
    \end{aligned}
    \end{equation}
    Here, $\sigma_{read}$ represents the standard deviation of the read noise in the detector. 
    
    Differentiable noise modeling is essential to update the forward model's physical parameters through back-propagation. Differentiability of the read noise is not a concern as it is content-independent. We derive differentiable Poisson sampling through the normal approximation to Poisson distribution \cite{Cheng1949TheNA} as shown in eq. \ref{eq:diff_pos_noise}
    
    \begin{equation}
    \begin{aligned}
    y^{poiss}_{down, t} & \sim \operatorname{Poiss}\left(\alpha_{down, t}\right) \\
    \Rightarrow y^{poiss}_{down, t} & \sim \mathcal{N} \left(\mu=\alpha_{down, t}, \sigma=\sqrt{\alpha}_{down, t}\right) \\
    \Rightarrow y^{poiss}_{down, t} &=\alpha_{down, t}+\sqrt{\alpha_{down, t}} \times z
    \end{aligned}
    \label{eq:diff_pos_noise}
    \end{equation}
    
    Here, $z \sim \mathcal{N}(\mu= 0,\sigma= 1)$. We added a $10$ photon background to $\alpha_{down, t}$ so that the minimal photon count received by the detector from the imaging system is larger enough to hold the approximation. This addition can be safely corrected for the real data by adding the same background to the real signal. In other words, the model always assumes a worse noise condition than the real-world case. Through the proposed method, we separate the random node from the computational graph hence making the stochastic noise model differentiable. This method is influenced by \textit{the reparameterization trick} from the work done by Kingma et al. \cite{vae2014}.

\begin{figure}[hbt!]
\begin{center}
\includegraphics[width=\linewidth]{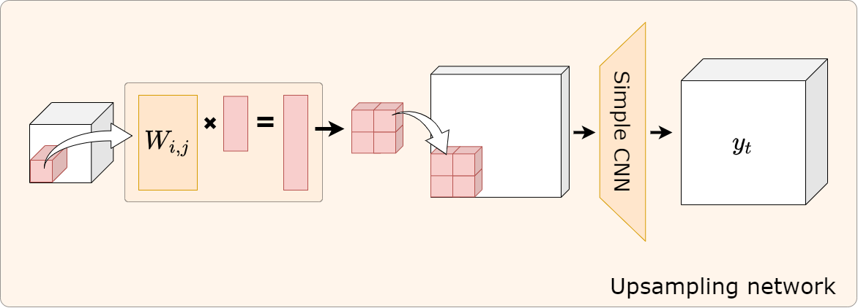}
\end{center}
\caption{Upsampling network: Each $(i, j)^{th}$ pixel in the acquired downscaled image $y_{down, t}$ is projected through weight matrix $W_{i, j}$. The resultant vectors are then reshaped into the upsampled $(i, j)^{th}$ patch. After tiling all the patches, the generated feature map then expanded to $y_t$ having $T$ channels. $T$ is the number of excitation patterns.}
\label{fig:upsample}
\end{figure}

\subsection{The Inverse Model}
\label{sec:inverse_model}

The inverse model reconstructs the sample from the acquired patterns through the imaging process. We propose a \textit{locality aware upsampling method} (Fig. \ref{fig:upsample}) along with a classical convolutional neural network-based reconstruction model for reconstruction from heavily compressed samples. 

In the proposed locality aware upsampling network, for each pixel location $(i, j)$ in the acquired demagnified image $y_{down, t}$, we define a learnable weight matrix $W_{i, j}$. The size of the weight matrix is determined by the upsampling factor. After $y_{down, t}(i, j)$ is projected using the weight matrix $W_{i, j}$, it is reshaped into a patch which is considered to be the upsampled patch from the pixel $y_{down, t}(i, j)$. Then a simple convolutional neural network with convolution, ReLU, and max-pooling layers converts the number of channels to the number of excitation patterns. The result is $y_t$. The architecture is motivated by the forward model operations where the patch details are encoded into a single pixel in $y_{down, t}$. Experimental results have shown that our proposed method induces a higher capability of unfolding compressed images hence resulting in better reconstructions at extreme compressions. 

We consider a conventional convolutional neural network for the reconstriction which consists of convolution, max-pooling, and ReLU layers with the final activation of Sigmoid. Due to the separation of the upsampling block from the reconstruction network, state-of-the-art super-resolution reconstruction pipelines can be utilized with the proposed upsampling block for higher-quality reconstructions (please refer to section \ref{swinir_results}).

More details regarding the architecture can be found in the supplementary.

\subsection{End-to-end Training}
\label{sec:end-to-end_training}

\subsubsection{Objective Function}

We utilize the conventional reconstruction objective function (eq. \ref{eq:objective}) as the default objective function for the comparisons unless otherwise specified.

\begin{equation}
    \label{eq:objective}
    \mathcal{L}_{L1}=\mathbb{E}_X\left[\|X_{recon}-X\|_{1}\right],
\end{equation}
where $X, X_{recon}$ represents the input image to the forward model and the reconstructed image. $\mathbb{E}_X[.]$ represents the expectation over the distribution of input images. 

\subsubsection{Convergence of the End-to-end Model}
\label{sec:convergence_problem}
The optical setup of the forward model typically works with the photons distributed in the range of 1000s. Deep neural networks/ gradient-based optimizations usually work better with data in the range of $[0, 1]$. This makes the end-to-end joint optimization of forward and inverse models unstable. The optical model cannot be normalized in a straight-forward way to match with the inverse model due to the addition of stochastic noise in the photo-detector model. Therefore we introduce a normalization method for the optical forward model to stabilize the end-to-end training of the proposed framework. The details can be found in the supplementary.

\subsubsection{Implementation Details}

All the implementations are done through the PyTorch deep learning framework with Python 3.6. Adam optimizer \cite{adam} is used to optimize the weights in both forward and inverse models. Default learning rates for the forward and inverse models are 1.0 and 0.001 respectively. Fast Fourier Transform \cite{rao2010_fft} is utilized to mimic the Fourier transform in the forward model. We used batch size 32 for all the experiments.

We train the content-aware experiments with the following procedure. First, we only train the inverse model until the validation loss gets plateaued. The number of epochs depends on the dataset (e.g. For the U2OS cell dataset in section \ref{sec:datasets}, we consider 12150 epochs). Then we start to optimize the illumination patterns along with the inverse model with the schedule presented in algorithm \ref{eq:m-schedule} (For the U2OS dataset, we train the end-to-end model for another 12150 epochs). 

After the training, we obtain the optimal excitation patterns from $H_t$ which are learned for particular data distribution and a task. At the test time, these learned excitation patterns are used to illuminate the samples. The learned inverse model reconstructs the image from the detections.

\section{Experiments and Results}

Using a number of datasets (section \ref{sec:datasets}), we used our $\partial\mu$: 1) on content-aware sampling (section \ref{sec:content-aware}), 2) on segmentation-aware sampling (section \ref{sec:task-aware}), 3) to further analyze the new upsampling network (section \ref{sec:upsampling-network}), 4) to analyze robustness to noise (section \ref{sec:robustness_to_noise}), and 5) on high-resolution image reconstruction with state-of-the-art super-resolution pipelines (SwinIR) (section \ref{swinir_results}).

\subsection{Experimental datasets}
\label{sec:datasets}

    \paragraph{PatchMNIST digits} We design PatchMNIST digits dataset to make the MNIST digits dataset \cite{deng2012mnist}  more complex. Here we first resize the MNIST images to $32 \times 32$ and then tile them to create a $20 \times 20$ image grid. $256 \times 256$ size patches are then extracted from the image grid to create the dataset. 3000 training images, 375 validation, and 375 test images are generated. For the validation and test images, only the test set of MNIST digits is used. Therefore the resultant validation and test images do not contain any image from the initial training set of MNIST data. We use this dataset to evaluate the performance of content-aware sampling, to show the superiority of the proposed upsampling method, and to show noise robustness.
    
    \paragraph{U2OS Cell Dataset}
    U2OS (bone osteosarcoma) cells are fixed with 4\% paraformaldehyde and stained with DAPI. The cells are then imaged using a  spinning disk confocal microscope at 63$\times$ magnification using an objective with 1.4 numerical aperture. The procedure results in image stacks of size $60$$\times$$2304$$\times$$2304$. 
    
    We first apply maximum intensity projection on each stack to obtain the dataset having images with $2304\times2304$ size. After reducing the camera bias of $134.28$ from the obtained images, we clip the intensity by 500 to remove outliers and applied min-max normalization \cite{patro2015normalization}. We then downscale the images by a factor of 63/ 20 and obtain the image size of $731 \times 731$. The resultant dataset is divided into the train, validation, and test sets with 168, 21, and 21 images.
    
    To train the models, we randomly cropped $256 \times 256$ patches from the train set images and applied random horizontal and vertical flips. To validate, and test the models, we considered $256 \times 256$ patches from the original validation, test sets.
    
    The dataset is used to evaluate the compressibility in content-aware sampling and task-aware (segmentation-aware) sampling.
    
    
    \paragraph{Human MCF7 cells} Human MCF7 cell dataset \cite{bbbcdataset} is an opensource dataset of MCF-7 breast cancer cells. Images from channel-2 of the dataset were utilized to show the generalizability of our method. The chosen channel contains images with visually different features compared to the other selected datasets. To keep the consistency among the experiments,  3000, 100, and 100 image patches were considered as the train, validation, and test sets respectively.

    \paragraph{Div2K, Flickr2K, Set5, Set14, BSD100, Urban100, Manga109} We utilized standard super-resolution datasets; Div2K \cite{div2k_dataset}, Flickr2K \cite{div2k_dataset} to train our method on classical grayscale super-resolution task. We evaluated the performance on standard super-resolution test datasets: Set5 \cite{set5_dataset}, Set14 \cite{set14_dataset}, BSD100 \cite{bsd100_dataset}, Urban100 \cite{urban100_dataset}, Manga109 \cite{manga109_dataset}.

\subsection{Content-aware sampling}
\label{sec:content-aware}

We test the hypothesis that \textit{content-aware sampling on the proposed microscope can achieve better compression compared to traditional compressive sampling techniques on the same system}. We compare traditional compressive sampling techniques including uniform (i.e. wide-field illumination), random (pseudo-random illumination), and Hadamard (patterns engineered using the Hadamard basis).

\begin{figure*}[hbt!]
\vspace{-0.2cm}
\begin{center}
\includegraphics[width=1.9\columnwidth]{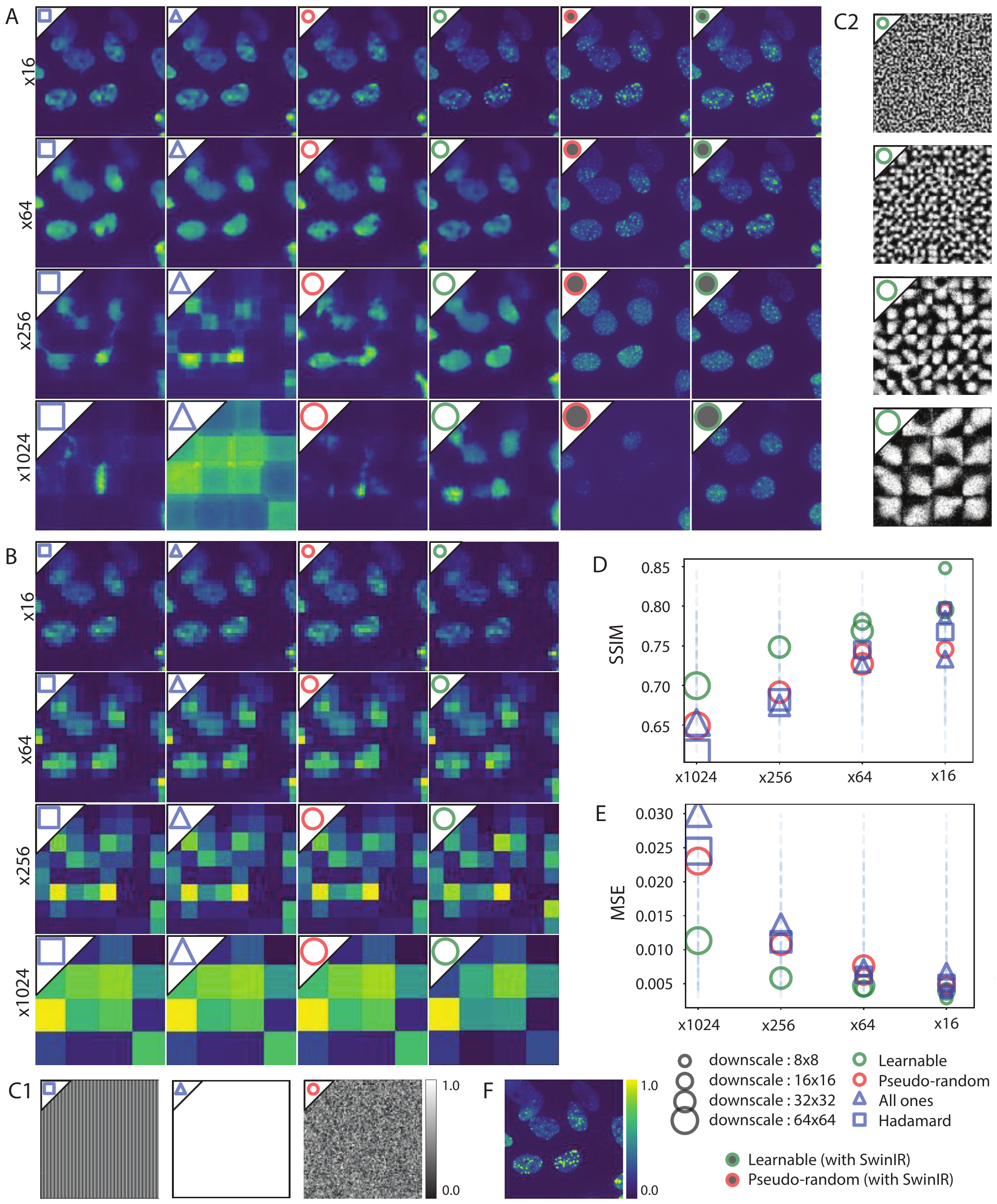} 
\end{center}
\caption{Results for content-aware reconstruction. A) Reconstructions for a random sample from the test set. Each row represents experiment results for different compression, Each column represents results for Hadamard, uniform, pseudo-random, learnable (our) illuminations, pseudo-random + SwinIR reconstructions, and learnable + SwinIR reconstructions (our) respectively. B) Corresponding detections before the reconstruction pipeline (normalized by the maximum value of the field). C1) Example illumination pattern for Hadamard, uniform, pseudo-random illuminations.  C2) Example learned illumination patterns (for $8 \times 8$, $16 \times 16$, $32 \times 32$, $64 \times 64$ downscaling). D, E) Corresponding SSIM, MSE results for the experiments. F) Ground truth test image. We consider $10000$ photon count for all the experiments here. Quantitative results for SwinIR reconstructions are presented in Table \ref{tab:quant-swinir}.}
\label{fig:main-res}
\end{figure*}

The qualitative performances on a representative test image are shown in Fig. \ref{fig:main-res}. Learned patterns (Fig. \ref{fig:main-res} (C1, C2)) suggest that learnable illumination enables the optical model to encode content-aware information in the illumination itself. Higher qualitative (Fig. \ref{fig:main-res} (A)) and quantitative (Fig. \ref{fig:main-res} (D, E)) performances show that this phenomenon allows higher compression at the detector.

\textbf{Explanation of the quantitative plots (Fig. \ref{fig:main-res} (D, E))}: Fig. \ref{fig:main-res} (D) show the test SSIM of the reconstructions through the proposed method (\textit{green circles}) and other sampling methods. Let's consider the upper plot where $\text{SSIM}$ is plotted against the compression. Here each symbol corresponds to one experiment. 4 shapes (square, triangle, green circle, red circle) represent the different sampling methods. For each method, multiple experiments were conducted for different downscaling factors (which are represented by the size of the symbol) and different numbers of excitation patterns. For a given experiment, the number of excitation patterns can be computed using the compression and the downscaling factor. Compression is defined as  $\frac{\# \text{total pixels}}{\# \text{total detections}} = \frac{\# \text{pixels in image}}{\# \text{pixels in detector} \times \# \text{patterns}} = \frac{\text{amount of dowscaling}}{\# \text{patterns}}$. Higher compression results in a lower number of total measurements therefore higher throughput. For experiments with compression= $\times 64$ and downscaling= $32 \times 32, 16 \times 16$, the number of patterns can be obtained as  $16, 4$ respectively ($64= \frac{32^2}{16} = \frac{16^2}{4}$).

\begin{figure*}[hbt!]
\vspace{-0.2cm}
\begin{center}
\includegraphics[width=2\columnwidth]{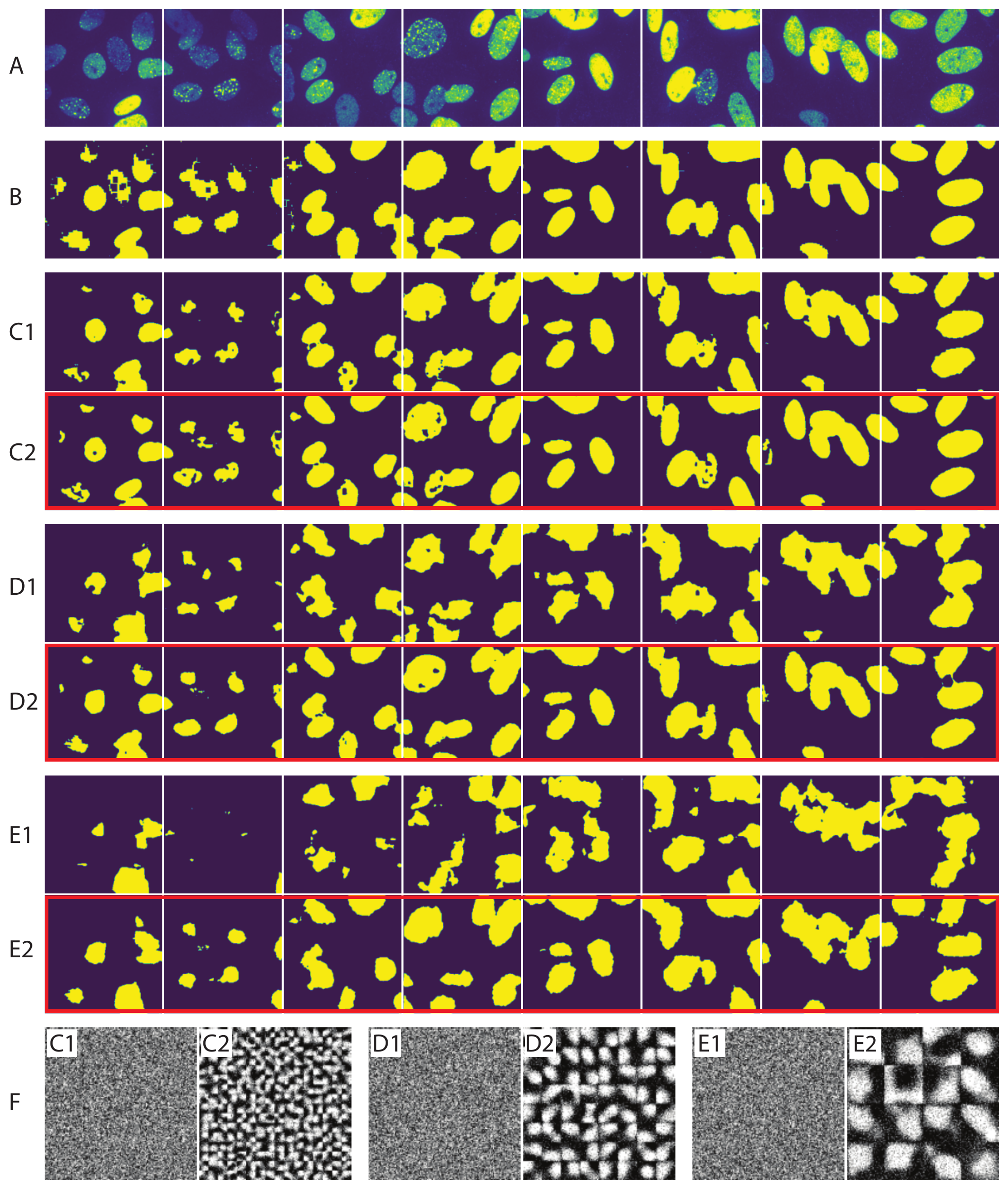} 
\end{center}
\caption{Segmentation-aware training. A) Ground truth images from the test set, B) Generated ground truth pseudo-segmentation maps, C1) Segmentation results with $\times 64$ compression, pseudo-random $H_t$, C2) Segmentation results with $\times 64$ compression, learnable $H_t$ (proposed), D1) Segmentation results with $\times 256$ compression, pseudo-random $H_t$, D2) Segmentation results with $\times 256$ compression, learnable $H_t$ (proposed), E1) Segmentation results with $\times 1024$ compression, pseudo-random $H_t$, E2) Segmentation results with $\times 1024$ compression, learnable $H_t$ (proposed), F) Representative illumination pattern for all C1- E2 experiments}
\label{fig:seg-main}
\end{figure*}




\subsection{Task-aware sampling}
\label{sec:task-aware}

In most of the imaging modalities, the obtained images are then used for particular tasks such as cancer diagnosis \cite{fass2008}, etc. But when compressively measured, the acquired image may lack features that are useful for that particular task while containing features that are not important for the task. 
 
To this end, we propose a task-aware configuration for proposed differentiable microscopy. The goal is to learn to sample the most important features of the image that are needed for the downstream task. For this demonstration, we picked \textit{segmentation} which is a common low-level task. We generated pseudo-ground truth segmentation maps to train the models according to the procedure in section \ref{sec:seg-gt}. Please find the training procedure in the supplementary (\ref{sec:seg-training}).

Fig. \ref{fig:seg-main} shows the segmentation results from pseudo-random and learnable (proposed) illuminations for $\times 64, 256, 1024$ compressions. The proposed method consistently generates better segmentation maps.






\subsection{Analysis of Proposed Upsampling Network}
\label{sec:upsampling-network}

\begin{figure}[hbt!]
\centering
\includegraphics[width=1.0\linewidth]{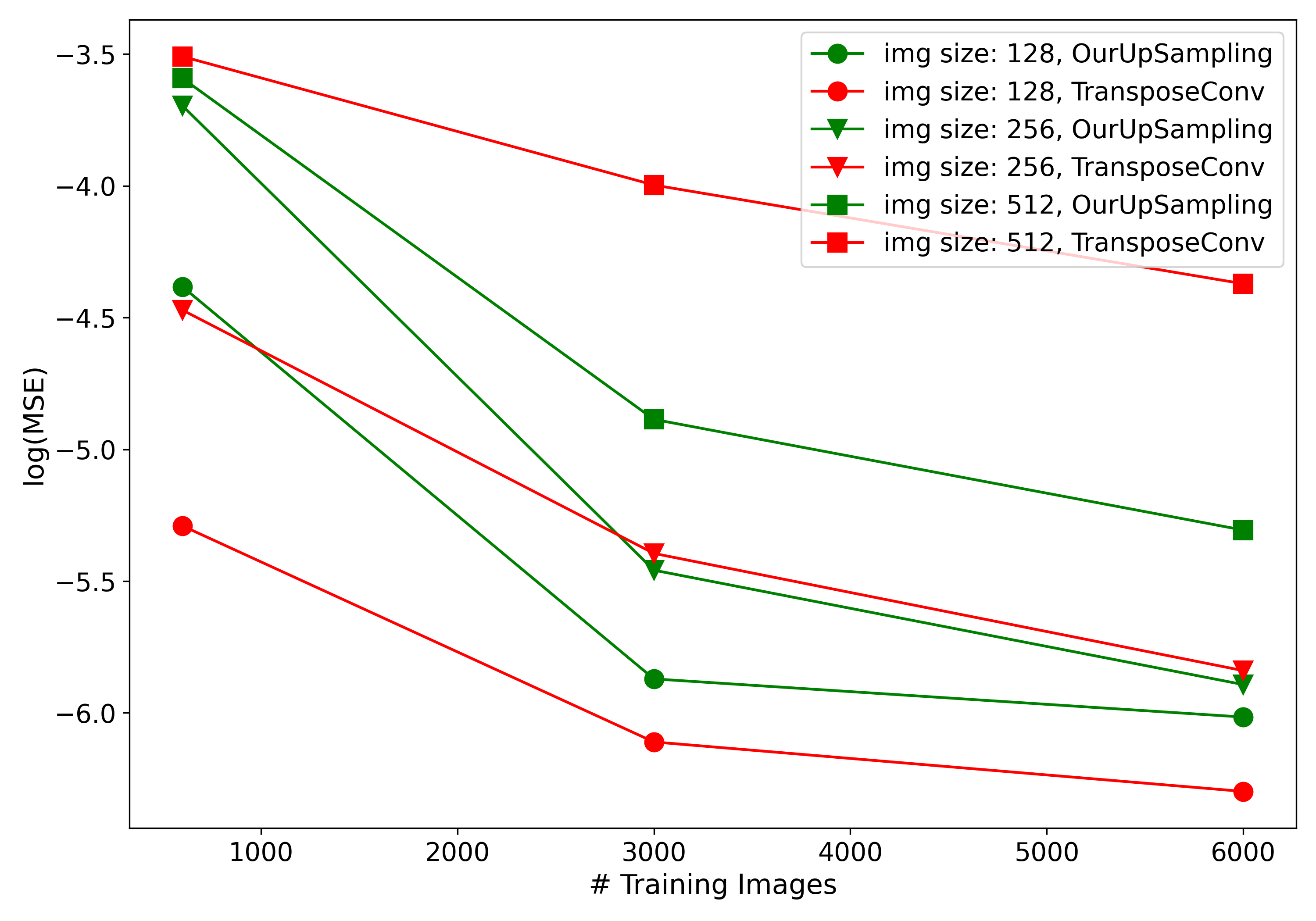}
\caption{$log(MSE)$ vs $\#$train images (on PatchMNIST dataset at $\times 8$ compression): Proposed upsampling network outperforms transpose convolution on large images specifically when there are sufficient training images. In contrast, transpose convolution fails to reconstruct large images even if there are sufficient amounts of training images.}

\label{fig:upsamp_n_samples}
\end{figure}

We evaluate the performance of the proposed upsampling network on different image sizes and different numbers of training images on the PatchMNIST dataset with $\times 8$ compression (Fig. \ref{fig:upsamp_n_samples}). The proposed upsampling network has less performance compared to transpose convolution when we trained it on a lower number of training images with smaller sizes. But when those two factors are getting higher, the proposed method outperforms transpose convolution by a considerable margin. This further tells that transpose convolution struggles to reconstruct larger images even when there is a sufficient number of training images. In contrast, the proposed network takes the advantage of the higher number of training images, therefore, performs better reconstruction.

\subsection{Robustness to Noise}
\label{sec:robustness_to_noise}

To illustrate the robustness of our method to noise, we consider extreme read noise and Poisson noise conditions. 

\begin{figure}[hbt!]
\centering
\includegraphics[width=1.0\linewidth]{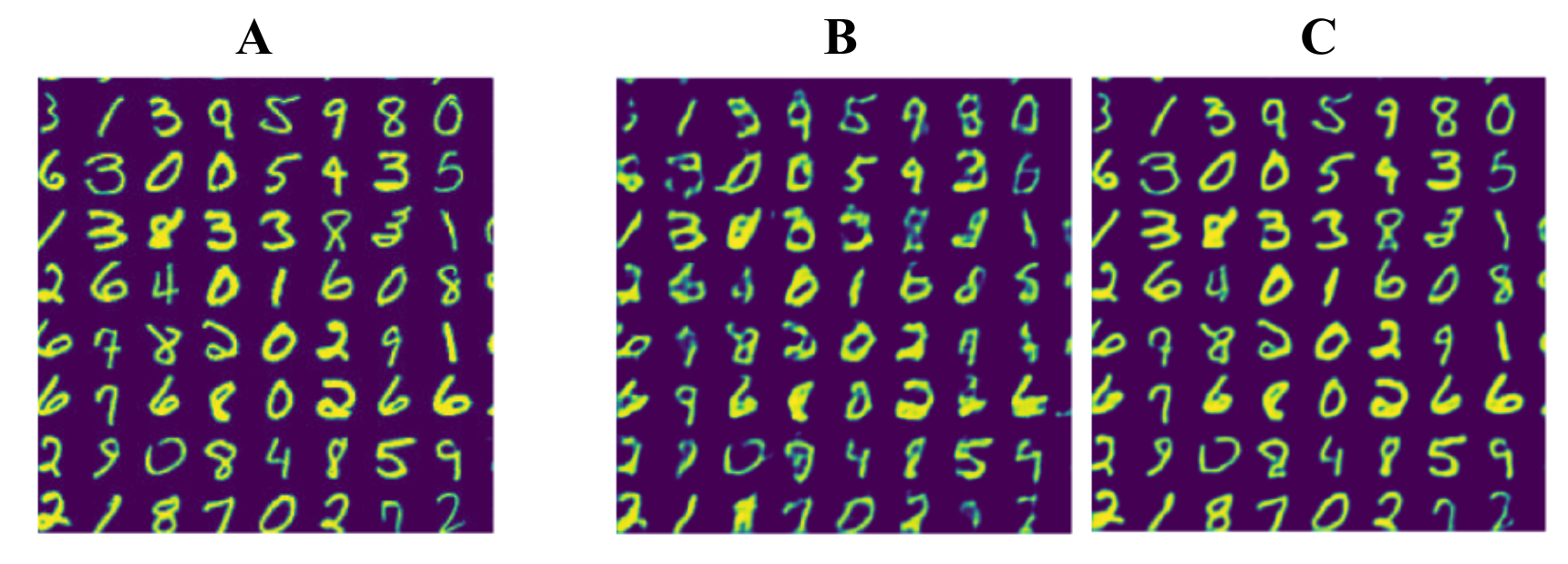}
\caption{Robustness to extreme noise condition with read noise standard deviation= 6.0 with photon count= 10.0: Ground truth image (A), Reconstructions from fixed random initialized excitation patterns (B), Reconstructions from our method (C)}
\label{fig:noise}
\end{figure}

Table \ref{fig:heatmap} shows the robustness of the method to different Poisson and read noise conditions. We observe that for each noise condition, the proposed method outperforms the fixed random pattern illumination while having consistent performance across each photon count regardless of the read noise. In Fig. \ref{fig:noise}, we demonstrate this performance improvement by considering the extreme noise conditions where the read noise standard deviation= 6.0 and the photon count= 10.0. As the figure suggests, the reconstructions from our method are closer to ground truth.

\begin{table}[hbt!]
\centering
\begin{tabular}{lcccc}
\hline
\multirow{1}{*}{$\sigma_{read}$} & \multicolumn{2}{c}{Our} & \multicolumn{2}{c}{\begin{tabular}[c]{@{}c@{}}Fixed Excitation\\ Patterns\end{tabular}} \\ \cline{2-5} 
 & \begin{tabular}[c]{@{}c@{}}photon\\ count\\ =10\end{tabular} & \begin{tabular}[c]{@{}c@{}}photon \\ count\\ = 10000\end{tabular} & \begin{tabular}[c]{@{}c@{}}photon \\ count\\ =10\end{tabular} & \begin{tabular}[c]{@{}c@{}}photon \\ count\\ = 10000\end{tabular} \\ \hline
0.0 & \textbf{0.0025} & \textbf{0.0059} & 0.0108 & 0.0214 \\
2.7 & \textbf{0.0024} & \textbf{0.0058} & 0.0107 & 0.0210 \\
2.0 & \textbf{0.0024} & \textbf{0.0061} & 0.0107 & 0.0213 \\
6.0 & \textbf{0.0024} & \textbf{0.0069} & 0.0107 & 0.0235 \\ \hline
\end{tabular}
\caption{Robustness to Poisson and Read Noise for PatchMNIST dataset: MSE of the reconstructions from the proposed method vs pseudo-random initialized illumination for $\times 8$ compression with T= 8}
\label{fig:heatmap}
\end{table}

\begin{table*}[hbt!]
\centering
\resizebox{\textwidth}{!}{
\begin{tabular}{lcccccccccc}
\multirow{2}{*}{Method} & \multicolumn{2}{c}{Set5} & \multicolumn{2}{c}{Set14} & \multicolumn{2}{c}{BSD100} & \multicolumn{2}{c}{Urban100} & \multicolumn{2}{c}{Manga109} \\ \cline{2-11} 
 & PSNR & SSIM & PSNR & SSIM & PSNR & SSIM & PSNR & SSIM & PSNR & SSIM \\ \hline
\begin{tabular}[c]{@{}l@{}}SwinIR w/o LI \end{tabular}& 14.03 & 0.3079 & 13.64 & 0.2258 & 14.28 & 0.2094 & 13.51 & 0.2146 & 12.09 & 0.1952 \\
\begin{tabular}[c]{@{}l@{}}SwinIR with LI \end{tabular} & \textbf{26.74} & \textbf{0.8113} & \textbf{23.6} & \textbf{0.693} & \textbf{22.9} & \textbf{0.6317} & \textbf{21.51} & \textbf{0.6402} & \textbf{20.18} & \textbf{0.6652} \\ \hline
\end{tabular}}
\caption{Evaluation of the proposed learnable illumination (LI) with SwinIR inverse model (for $\times 16$ compression)}
\label{tab:swinIR_general}
\end{table*}

\begin{figure*}[hbt!]
\centering
\includegraphics[width=1.0\linewidth]{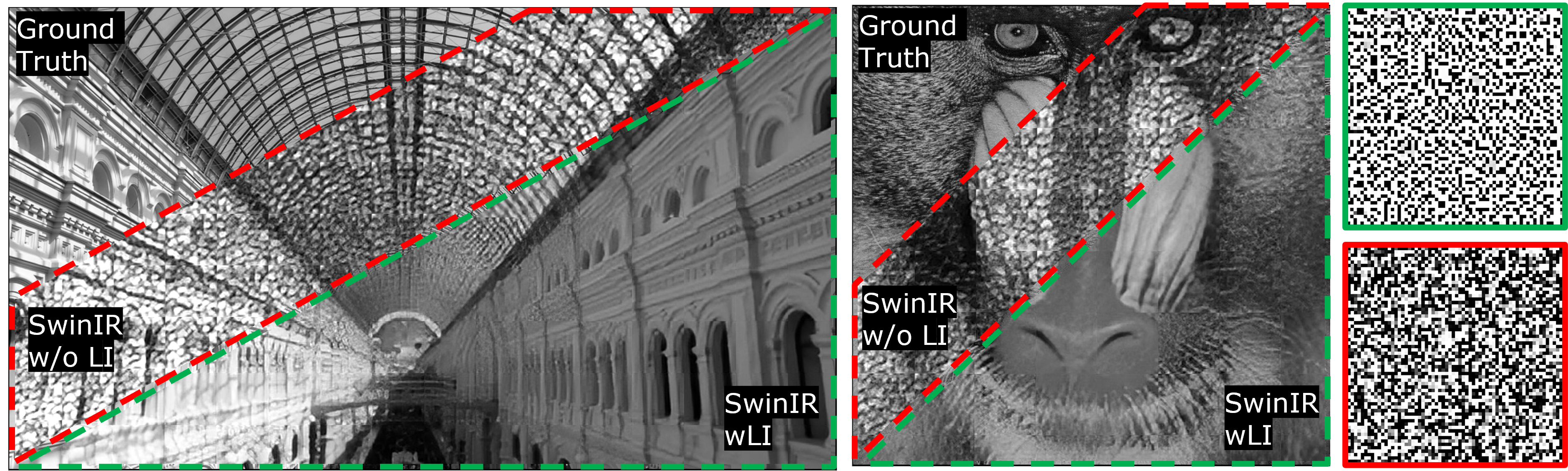}
\caption{Qualitative SwinIR based high-quality reconstructions on standard super-resolution datasets (Table \ref{tab:swinIR_general}) with $\times 16$ compression. Corresponding illumination patterns ($H_t$) are presented on the right side.}
\label{fig:swin_ir_standard_qualitative}
\end{figure*}

\begin{figure}[hbt!]
\centering
\includegraphics[width=1.0\linewidth]{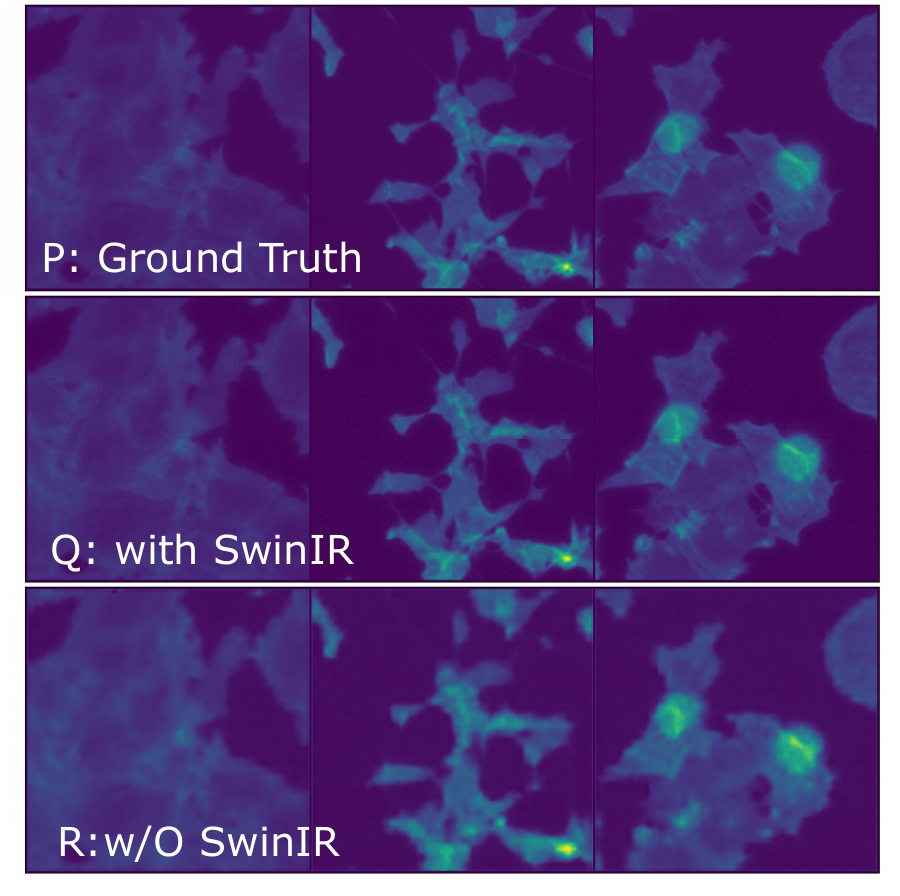}
\caption{Qualitative results from SwinIR reconstruction model (Q), simple transpose convolution reconstruction model (R) for $\times 16$ compression on HumanMCF7 cells.}
\label{fig:swinir_cell}
\end{figure}

\begin{figure*}[hbt!]
\centering
\includegraphics[width=1\linewidth]{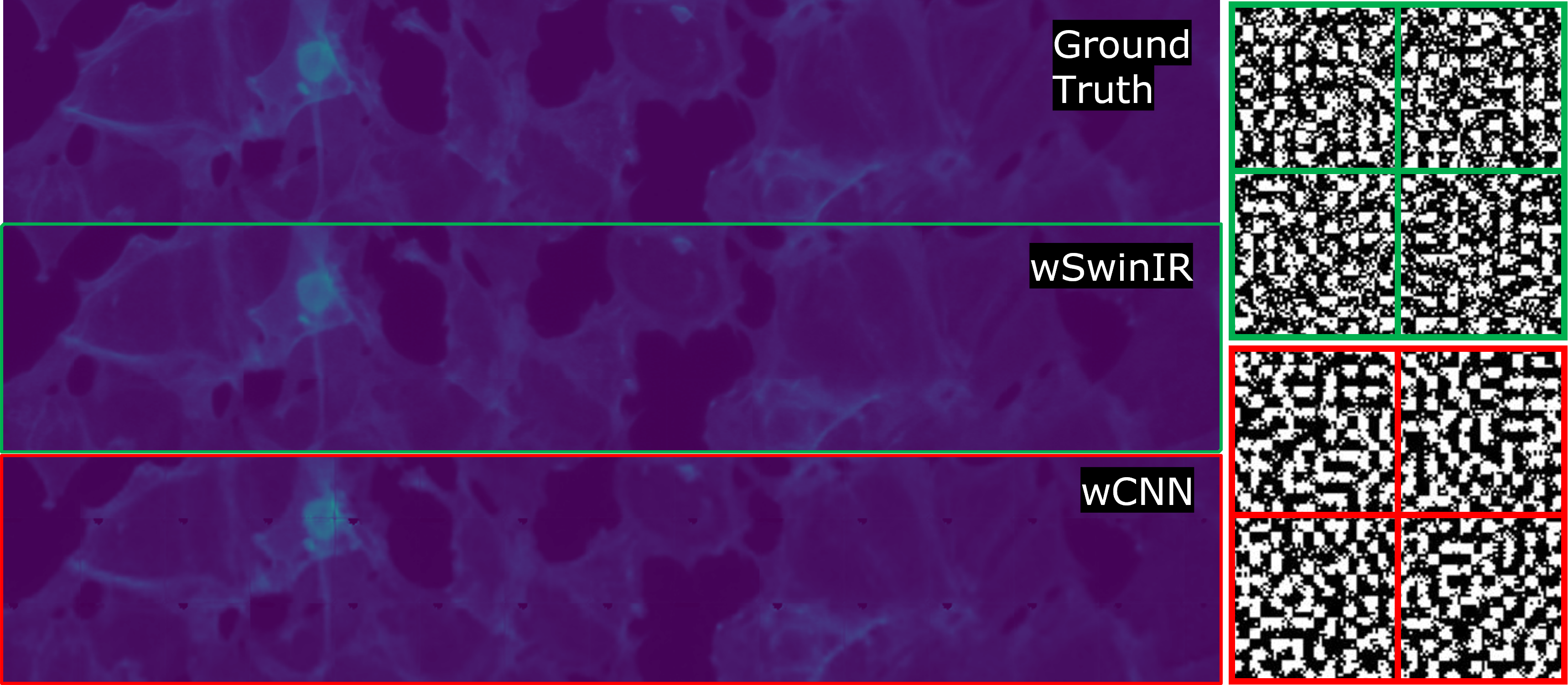}
\caption{More SwinIR based high-quality reconstructions on HumanMCF7 cells for $\times 16$ compression. In contrast to Fig. \ref{fig:swinir_cell} (where we present the improvement due to SwinIR as the reconstruction method compared to our conventional pipeline), here we include the results from CNN-based reconstruction with $64 \times 64$ image size for a fairer comparison. SwinIR-based reconstruction contains fewer border artifacts and higher visual image quality. Corresponding learned illumination patterns are included on the right}
\label{fig:swinir_cell2}
\end{figure*}

\subsection{High-resolution image reconstruction using SwinIR}
\label{swinir_results}

We replace the reconstruction model ($\psi$) in Fig. \ref{fig:end2end} from the state-of-the-art super-resolution network SwinIR \cite{swinir}. Similar to SwinIR real-world image super-resolution task, we utilize pixel loss, adversarial loss, and perceptual loss to train the network end-to-end. We use the initial learning rate of 0.1 for $H_t$. All other configurations are similar to SwinIR configurations \cite{swinir}. 

We first demonstrate that we can perform more sophisticated compression through learnable illumination. We train SwinIR on Div2K, and Flickr2K datasets with and without learnable pattern illumination. The training is performed for $64 \times 64$ image patches with $\times 16$ compression. The testing is performed on standard super-resolution test datasets in section \ref{sec:datasets}. We achieve up to 12, 0.51 PSNR, SSIM improvement compared to SwinIR without learnable illumination (Table \ref{tab:swinIR_general}). Qualitative results are presented in Fig. \ref{fig:swin_ir_standard_qualitative}.

We conduct a similar set of experiments with the U2OS cell dataset. We first train the proposed content-aware algorithm with and without learnable $H_t$. We append the SwinIR model at the end of the trained end-to-end model. Here we set the upscaling factor of the SwinIR to 1 (i.e. no upscaling). Finally, we train the SwinIR to super-resolve the output of the trained end-to-end model. The qualitative results are shown in Fig. \ref{fig:main-res}, \ref{fig:more_test}, \ref{fig:more_test_swinir}, quantitative results are shown in Table \ref{tab:quant-swinir}. We show that, with the proposed learnable $H_t$, SwinIR gives better reconstructions even at very high compressions.

Secondly, we show that the proposed realistic generalizable optical forward model along with the locality-aware upsampling block can be fused with any other super-resolution methods and objective functions. We use SwinIR as an image-to-image translation network without upsampling. The proposed upsampling network upsamples the images. Qualitative results for $256 \times 256$ reconstructions on Human MCF7 cells are shown in Fig. \ref{fig:swinir_cell}. Here we do not largely experiment with hyper-parameters to get the best results for this section since our goal is to show the applicability of the proposed method with other super-resolution pipelines. For a fairer comparison, we further compare the results with $64 \times 64$ image patch training without using SwinIR but using the classical convolutional reconstruction model explained in section \ref{sec:inverse_model}. As results in Fig. \ref{fig:swinir_cell2} suggest, the SwinIR-based method gives fewer border artifacts therefore superior reconstruction. To improve the stability of SwinIR-based training corresponding to Fig. \ref{fig:swinir_cell}, \ref{fig:swinir_cell2}, we utilize algorithm \ref{eq:m-schedule} with $epochs = 230, epoch step = 20, epoch_{cutoff}= 150, epoch_{baseline}= 150$.


\subsection{Ablation study}
\label{ablation_study}

\begin{figure}[hbt!]
\centering
\includegraphics[width=1.0\linewidth]{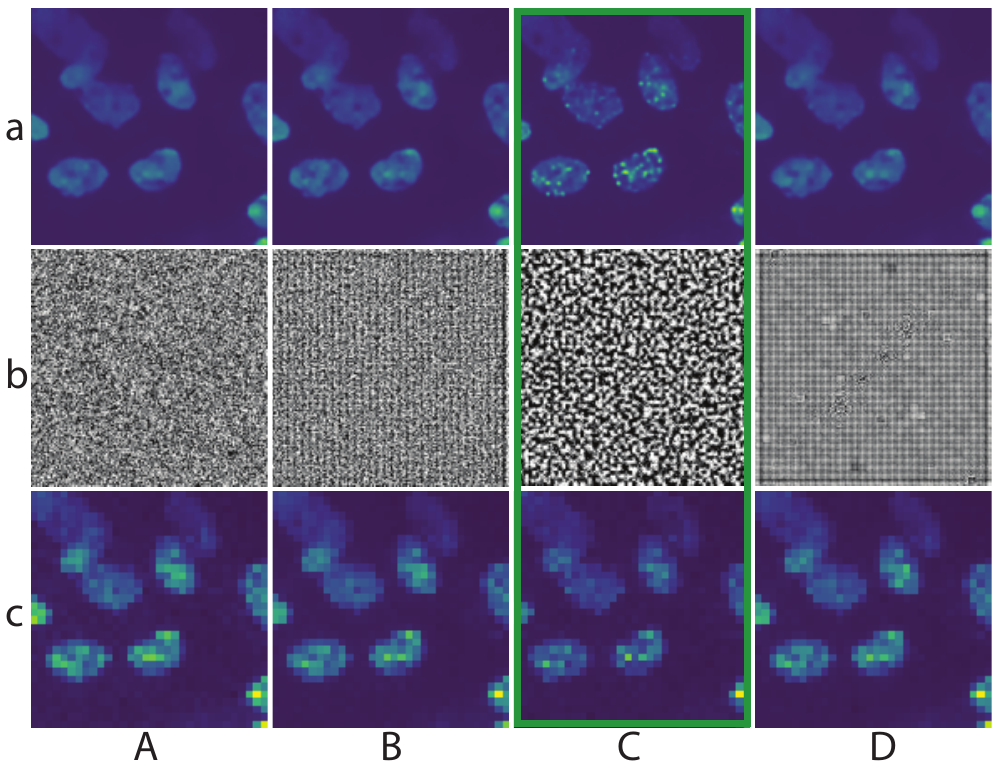}
\caption{Ablation study on U2OS cells at $\times 16$ compression with $T= 4$ and downscaling of $8\times8:1$ (qualitative results): A) fixed $H_t$ +  Tr.Conv.Up + frequency domain optimization. B) learnable $H_t$ +  Tr.Conv.Up + frequency domain optimization C) learnable $H_t$ + proposed upsampling + frequency domain optimization. D)  learnable $H_t$ + proposed upsampling + no frequency domain optimization. \textit{Tr.Conv.Up.} represents conventional transpose convolution upsampling.}
\label{fig:ablation}
\end{figure}

Table \ref{tab:ablation} shows the ablation study. Fig. \ref{fig:ablation} contains the corresponding qualitative results. The proposed method outperforms the baseline with a significant quantitative and qualitative improvement at $\times 16$ compression. 

Transpose Convolution learns generalized filters for images. Therefore they fail to reconstruct finer features. In contrast, the proposed upsampling block can learn locality-aware mappings while decoding much better connectivity of pattern pixels to detector pixels.

Furthermore, we conclude that frequency-domain optimization is important for training. Frequency domain conversion essentially results in convolution in the spatial domain, and convolutions capture objects irrespective of spatial locations. Since objects (eg: cells) might appear in any part of the field of view, we argue that spatial invariance is important for better performance.

\begin{table}[hbt!]
\centering
\begin{tabular}{lcc}
\multirow{1}{*}{Method} & \multicolumn{2}{c}{Performance} \\ \cline{2-3} 
 & \multicolumn{1}{c}{SSIM $\uparrow$} & \multicolumn{1}{c}{MSE $\downarrow$} \\ \midrule[1pt]
A: Baseline (fixed $H_t$ +  Tr.Conv.Up.)& 0.7872 & 0.0042 \\
B: (+) learnable Ht &  0.7950 & 0.0038 \\ 
C: (+) proposed upsampling network & \textbf{0.8426} & \textbf{0.0029}\\ 
D: (-) frequency domain optimization & 0.7857 & 0.0041 \\
\midrule[1pt]
\end{tabular}
\caption{Ablation study on U2OS cells at $\times 16$ compression with $T= 4$ and downscaling of $8\times8:1$ (quantitative results). \textit{Tr.Conv.Up.} represents conventional transpose convolution upsampling.} 
\label{tab:ablation}
\end{table}

\section{Limitations}

The major limitation of the proposed method is, the compression factor is not generic and depends on the compressibility inherited from the data distribution. This limitation exists in any sort of compression system. However, the proposed method allows different ways to identify, and partially/ completely overcome this issue. As an example, if identifiable important features are missing in the reconstructed images, one can 1. reduce the compression level until it gives reasonable performance. At this point, the compression level is similar to the compressibility limit inherited from the dataset or, 2. perform task-aware compression considering the specific task. This allows much higher compression because the model can neglect the redundant information which is not useful for the specific task.

\section{Conclusion}

In microscopy throughput and image quality are competing requirements. One is often traded off for the other in biological experiments. Many previous methods tried to solve this problem by post-processing already acquired images (especially using deep learning) or by traditional compressive sampling. The performances of such methods are bounded by the fixed optical configuration of the microscope. Few previous works did optimize the optical hardware of the microscope together with the reconstruction; They all focused on improving the image quality, not the throughput. Most were also limited to LED illumination sources, and simple \textit{iid} noise models. In this work, we propose more general end-to-end differentiable compressive microscopy. Our method consists of a realistic forward model for compressive acquisition and a robust inverse model for reconstruction. The forward model contains a realistic stochastic noise model and an excitation pattern optimization network. The reconstruction network contains a physics-motivated locality-aware upsampling method to unravel heavily compressed images. Our differentiable compressive microscopy outperforms traditional compressive sampling schemes on content-aware sampling. Furthermore, we propose a task-aware configuration for segmentation. Our configuration outperforms the competition in generating segmentation maps under heavy compressions. To the best of our knowledge, we are the first to incorporate \textit{deep learning for compressive sampling} through optimizing the optical forward model for microscopy.


{
    \small
    \bibliographystyle{ieee_fullname}
    \bibliography{egbib}
}

\clearpage

\appendix

\setcounter{page}{1}

\renewcommand\thefigure{\arabic{figure}}
\setcounter{figure}{0}
\setcounter{equation}{0}
\setcounter{table}{0}

\renewcommand{\thefigure}{S\arabic{figure}}
\renewcommand{\thetable}{S\arabic{table}}
\renewcommand{\theequation}{S\arabic{equation}}

\twocolumn[
\centering
\Large
\textbf{Differentiable Microscopy for Content and Task
Aware Compressive Fluorescence Imaging} \\
\vspace{0.5em}Supplementary Material \\
\vspace{1.0em}
] 

\appendix

\section{Further Details of the Method}

\subsubsection{Schedule Selection for Custom Sigmoid Hyper-parameters}

We found that for larger $epoch step$ in algorithm \ref{eq:m-schedule}, the end-to-end training gets unstable and diverges after passing the $epoch_{cutoff}$. Similarly, finding a better value for $epoch_{cutoff}$ also requires careful experimentation because, larger $epoch_{cutoff}$ might results in values for $H_t$ that are not near to $ \{0, 1\}$ while smaller $epoch_{cutoff}$ disturb the model convergence. We found that, for the U2OS cell dataset, the hyper-parameters $epoch_{cutoff}= 18630, epoch step= 810$  can give enough epochs for the model to converge to a better point while having further epochs to make the values of $H_t$ towards $\{0, 1\}$ without getting the models diverged. However, these parameters heavily depend on the size and complexity of the training set.

\subsection{Inverse model}
The inverse model consists of 2 networks namely \textit{upsampling network} and \textit{reconstruction network}. For the initial comparisons, we consider a conventional convolutional neural network for the reconstruction network. This has 6 \textit{conv\_relu\_bn} blocks that progressively convert the initial number of channels $T$ to $1$. Here \textit{conv\_relu\_bn} block contains cascaded convolution (kernel size= 3), ReLU, and 2d batch normalization layers except the last block where the last block contains convolution and a Sigmoid layer.

\subsection{Convergence of end-to-end model}

As explained in section \ref{sec:convergence_problem}, the end-to-end model with forward and inverse models suffers from convergence issues due to their operating range mismatch. Direct normalization of the forward model is erroneous because it changes the physical model's stochastic non-linear noise statistics. Therefore we propose a normalizing pipeline in the following sections for the forward model.

\subsubsection{Recalling the initial noise model with additional details}

Repeating the differentiable photodetector model in section \ref{sec:detector_model},

\begin{equation}
y_{down, t} =  y^{poiss}_{down, t} + y^{normal}_{down, t} 
\label{eg:pos_noise_suppl}
\end{equation}
where,

\begin{equation}
\begin{aligned}        
    y^{poiss}_{down, t} & \sim \operatorname{Poiss}(\alpha_{down, t})\\
    y^{normal}_{down, t} & \sim \mathcal{N}(\mu= 0, \sigma= \sigma _{read})
\end{aligned}
\label{eq:both_noise_suppl}
\end{equation}

$y^{poiss}_{down, t}$ can be further simplified as,
\begin{equation}
\begin{aligned}
y^{poiss}_{down, t} & \sim \operatorname{Poiss}\left(\alpha_{down, t}\right) \\
\Rightarrow y^{poiss}_{down, t} & \sim \mathcal{N} \left(\mu=\alpha_{down, t}, \sigma=\sqrt{\alpha}_{down, t}\right) \\
\Rightarrow y^{poiss}_{down, t} &=\alpha_{down, t}+\sqrt{\alpha_{down, t}} \times z
\end{aligned}
\label{eq:diff_pos_noise_suppl}
\end{equation}

\begin{figure*}[]
\centering
\includegraphics[width=1\linewidth]{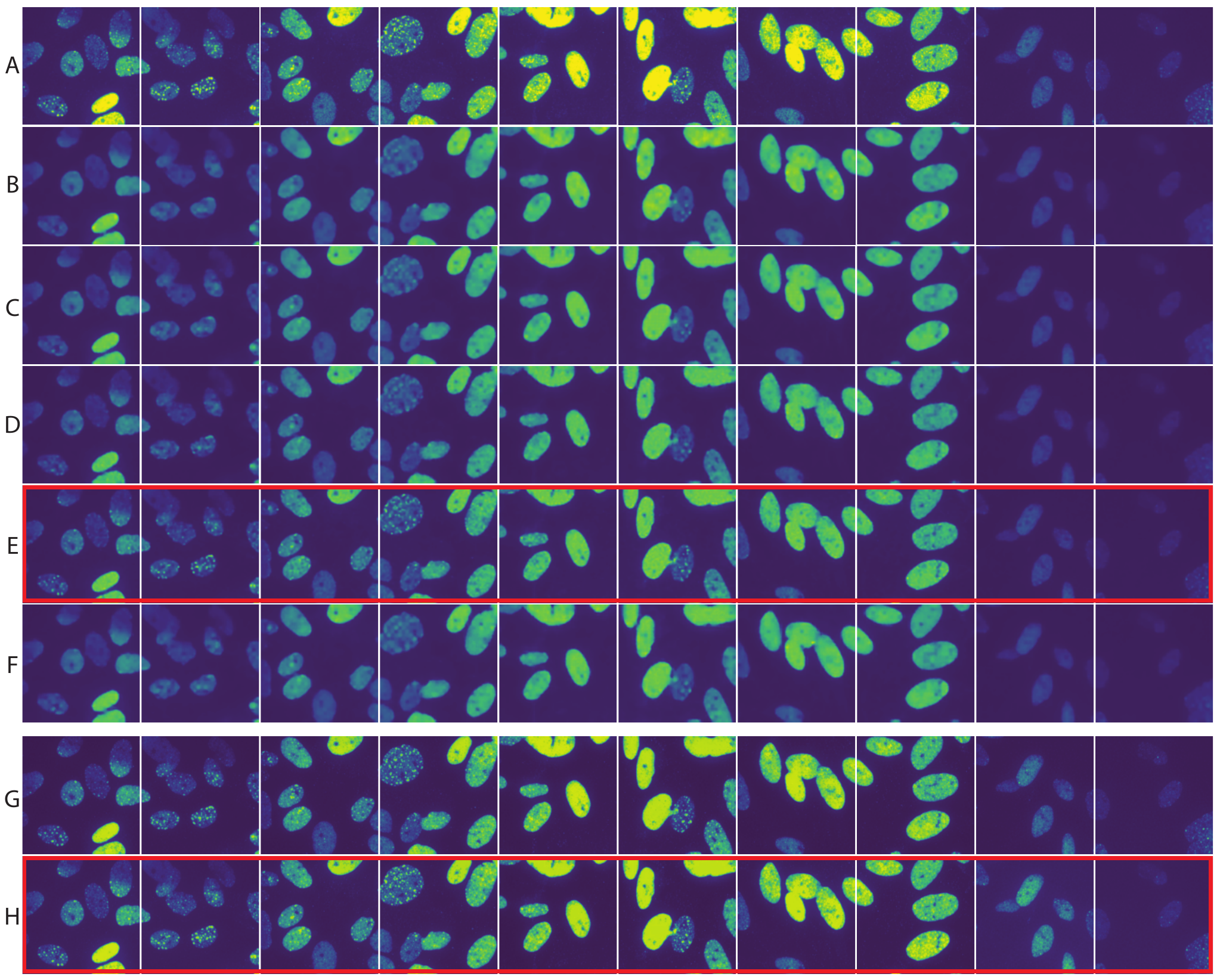}
\caption{Further test results of \textit{content aware} microscopy ($\times 16$ compression) on U2OS cell datset. A) Ground truth images from the test set, B) Reconstruction from baseline (\textit{fixed pseudo-random $H_t$} + \textit{Tr.Conv.Up} + Frequency domain optimization), C) learnable $H_t$ + \textit{Tr.Conv.Up} + Frequency domain optimization, D) fixed pseudo-random $H_t$ + proposed upsampling + Frequency domain optimization, E) learnable $H_t$ + proposed upsampling + Frequency domain optimization (Proposed method), F) learnable $H_t$ + Tr.Conv.Up. (\textit{without frequency domain optimization}), G) \textit{fixed pseudo-random $H_t$} + proposed upsampling + Frequency domain optimization + SwinIR super-resolution ,H) learnable $H_t$ + proposed upsampling + Frequency domain optimization + SwinIR super-resolution (Proposed method)}
\label{fig:more_test}
\end{figure*}

\begin{figure*}[]
\centering
\includegraphics[width=1\linewidth]{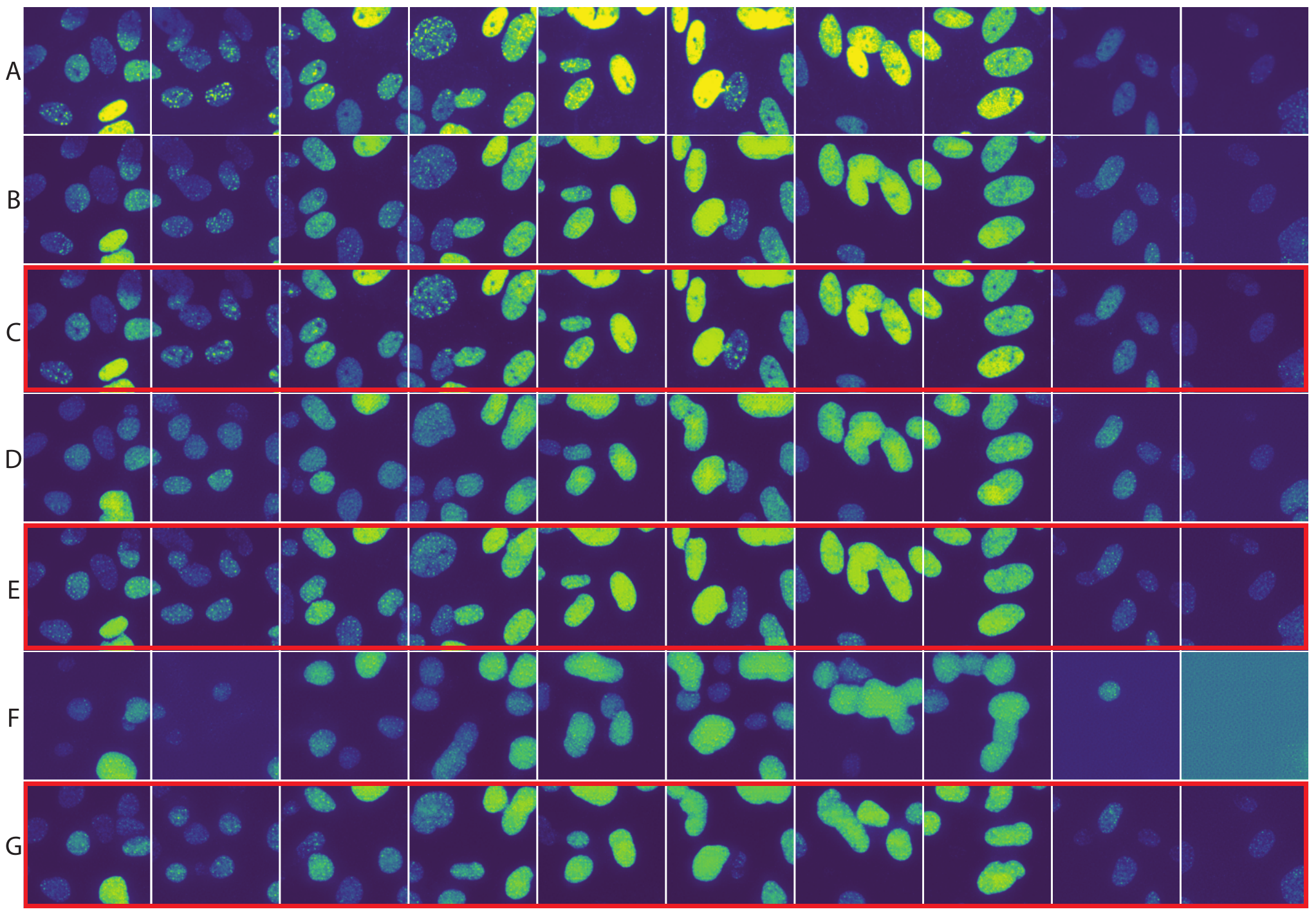}
\caption{Further test results of \textit{Content aware} microscopy with SwinIR reconstruction on U2OS cell dataset: A) Ground truth images from the test set, B) fixed pseudo-random $H_t$ + $\times 64$ comp., C) learnable $H_t$ + $\times 64$ comp. (proposed), D) fixed pseudo-random $H_t$ + $\times 256$ comp., E) learnable $H_t$ + $\times 256$ comp. (proposed), F) fixed pseudo-random $H_t$ + $\times 1024$ comp., G) learnable $H_t$ + $\times 1024$ comp. (proposed)}
\label{fig:more_test_swinir}
\end{figure*}

\subsubsection{Normalization of forward model}
\label{sec:norm_noise_model}

Normalization of the forward model is essential for better convergence of the end-to-end model due to the operating range mismatch between forward and inverse models. 

Consider $H_t \in [0, k]$ where $k$ is the highest photon count level of the excitation pattern. We can model the forward model before the detector through a linear function $\psi$. 

\begin{equation}
    \alpha_{down, t} = \psi(X, H_t)
\end{equation}

After incorporating $k$ with the linearity of the $\psi$,

\begin{equation}
\begin{aligned}
    \alpha_{down, t} &= k \psi(X, H_{t}^{norm})\\
    \alpha_{down, t} &= k \alpha_{down, t}^{norm}
\end{aligned}
\end{equation}

Substituting to eq. \ref{eq:diff_pos_noise_suppl}, 

 \begin{equation}
 \begin{aligned}
    y^{poiss}_{down, t}& = k \alpha_{down, t}^{norm} + \gamma \\&+ \sqrt{k \alpha_{down, t}^{norm} + \gamma} \times z
\end{aligned}
\end{equation}
where $z \sim \mathcal{N}(\mu = 0,\sigma = 1)$. We introduce $\gamma= 10$, an offset to make the normal approximation to Poisson valid.

Let $y^{poiss}_{down, t} = k\times y_{down, t}^{poiss, norm}$,

\begin{equation}
\begin{aligned}
    k y_{down, t}^{poiss, norm} &= k \alpha_{down, t}^{norm} + \gamma + \sqrt{k \alpha_{down, t}^{norm} + \gamma} \times z \\
    \Rightarrow    y_{down, t}^{poiss, norm} &= \alpha_{down, t}^{norm} + \frac{\gamma}{k}
    \\&\quad  + \sqrt{\frac{\alpha_{down, t}^{norm}}{k} + \frac{\gamma}{k^2}} \times z
\end{aligned}
\label{eq:normalized_poisson_noise}
\end{equation}

The read noise in eq. \ref{eq:both_noise_suppl} can be written as follows, 

\begin{equation}
y^{normal}_{down, t} =  \sigma_{read}z
\end{equation}
where, $z \sim \mathcal{N}(\mu=0, \sigma=1)$. The normalized read noise $y_{down, t}^{read, norm}$ is then obtained as in the eq. \ref{eq:normalized_read_noise}.

\begin{equation}
\begin{aligned}
y^{normal}_{down, t} &=  k\sigma^{norm}_{read}z\\
\Rightarrow y^{normal}_{down, t} &= ky_{down, t}^{read, norm}
\\
\Rightarrow y_{down, t}^{read, norm} &= \frac{y^{normal}_{down, t}}{k}
\end{aligned}
\label{eq:normalized_read_noise}
\end{equation}

The final normalized noise model is written as,

\begin{equation}
y^{norm}_{down, t} =  y^{poiss, norm}_{down, t} + y^{read, norm}_{down, t} 
\label{eq:final_pos_noise}
\end{equation}
where $y^{poiss, norm}_{down, t}, y^{read, norm}_{down, t}$ are obtained through eq. \ref{eq:normalized_poisson_noise}, \ref{eq:normalized_read_noise} respectively. The resulting $y^{norm}_{down, t}$ is then fed into the inverse model explained in the section \ref{sec:inverse_model} instead of $y_{down, t}$.

\subsubsection{Pseudo-ground Truths for Segmentation-aware Microscopy}
\label{sec:seg-gt}
To conduct segmentation-aware experiments, we created pseudo-ground truth segmentation maps for the U2OS cell dataset. First, we normalized the images into the range [0, 1]. Second, we apply thresholding to obtain a rough segmentation map. We experimentally found that a 0.3 threshold is good for the dataset. Finally, we applied \textit{closing} morphological operation with (10, 10) kernel to obtain the pseudo-ground truth segmentation maps.

\section{Experiment Details}



\subsubsection{Segmentation-aware microscopy: Training procedure}
\label{sec:seg-training}

Similar to content-aware microscopy, we consider a supervised learning framework to train segmentation-aware microscopy. Here the ground truths are the pseudo-ground truths we obtained in section \ref{sec:seg-gt}.

We train the end-to-end differentiable microscopy using the following steps. First, we train the end-to-end model for content-aware with the procedure described in Fig. \ref{fig:end2end}. This allows the illumination patterns and reconstruction network to learn about the data distribution more generally. Then we append and train a small convolutional network for the segmentation task. Here, all the parameters learned from the content-aware step were fixed. Finally, we finetune all the components (i.e. excitation pattern optimization network, inverse model, segmentation network) end-to-end. 

\section{Further results}

\begin{figure}[]
\centering
\includegraphics[width=1.0\linewidth]{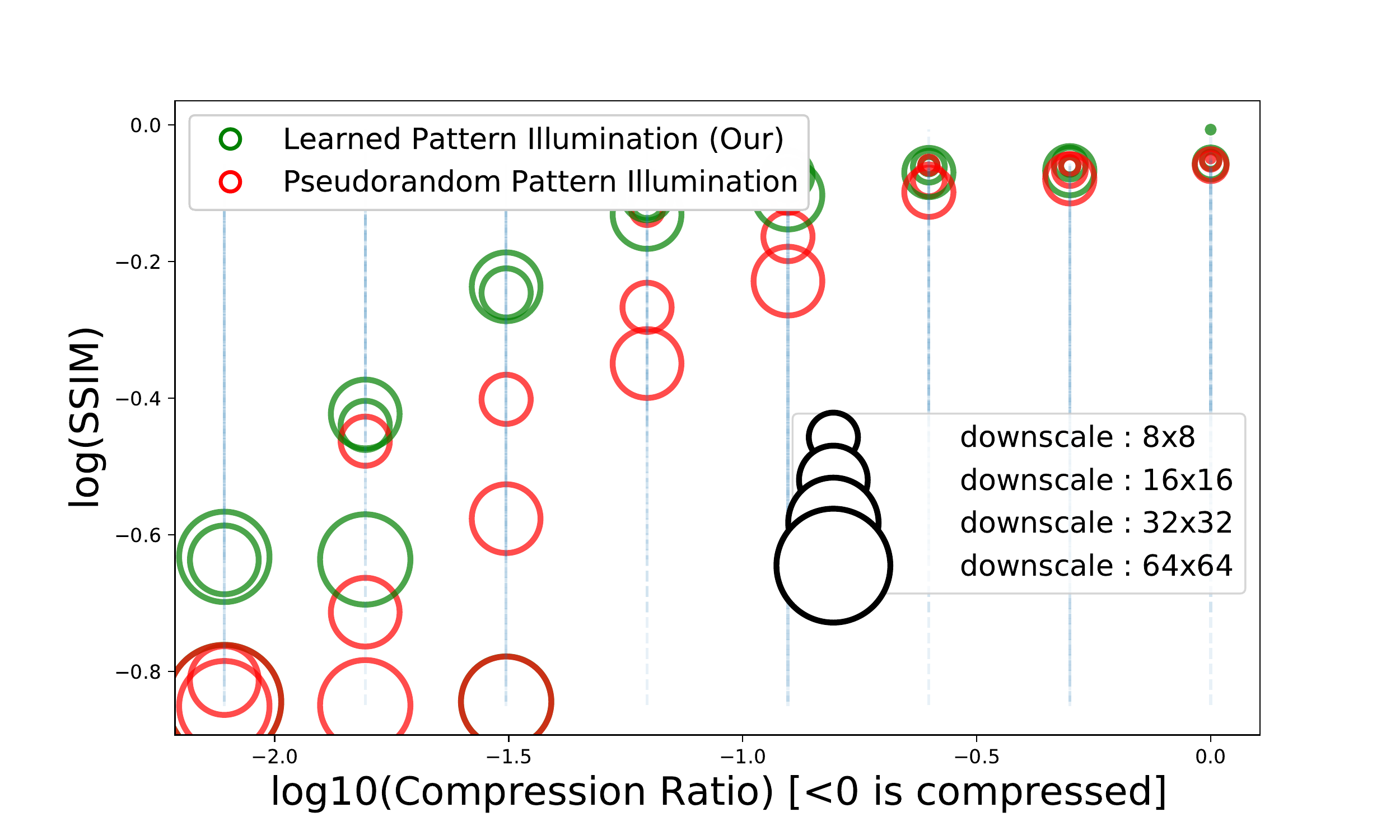}
\caption{Quantitative evaluation (log SSIM) of proposed content-aware microscopy on PatchMNIST dataset (photon count= 10000)}
\label{fig:quant_content_patchmnist}
\end{figure}

Fig. \ref{fig:more_test} shows more test results for content-aware sampling for $\times 16$ compression, Fig. \ref{fig:more_test_swinir} shows more test results for SwinIR based content-aware sampling for different compressions (Table \ref{tab:quant-swinir} contains corresponding quantitative results), and Fig. \ref{fig:quant_content_patchmnist} demonstrates quantitative results for content-aware sampling on PatchMNIST dataset.

\begin{table}[hbt!]
\centering
\begin{tabular}{lcc}

\multirow{1}{*}{Method} & \multicolumn{2}{c}{Performance} \\ \cline{2-3} 
 & \multicolumn{1}{c}{SSIM $\uparrow$} & \multicolumn{1}{c}{MSE $\downarrow$} \\ \midrule[1pt]

\textbf{$\times 16$ comp.}: pseudo-random & 0.7647 & 0.0038 \\ 
\textbf{$\times 16$ comp.}: learnable & \textbf{0.8174} & \textbf{0.0022} \\
\midrule[1pt]

\textbf{$\times 64$ comp.}: pseudo-random & 0.7146 & 0.0063 \\ 
\textbf{$\times 64$ comp.}: learnable & \textbf{0.7524} & \textbf{0.0040} \\
\midrule[1pt]

\textbf{$\times 256$ comp.}: pseudo-random & 0.6846 & 0.0113 \\ 
\textbf{$\times 256$ comp.}: learnable & \textbf{0.7149} & \textbf{0.0065} \\
\midrule[1pt]

\textbf{$\times 1024$ comp.}: pseudo-random & 0.6435 & 0.0259 \\ 
\textbf{$\times 1024$ comp.}: learnable & \textbf{0.6796} & \textbf{0.0133} \\
\midrule[1pt]
\end{tabular}

\caption{Quantitative results for SwinIR super-resolution experiments presented in Fig. \ref{fig:main-res}, \ref{fig:more_test_swinir}} 
\label{tab:quant-swinir}
\end{table}


\end{document}